\documentclass{article}
\usepackage{spconf,amsmath,graphicx}

\usepackage{mathtools}
\usepackage{hyperref}

\usepackage{caption}
\usepackage{subcaption}
\usepackage{booktabs}
\usepackage{adjustbox}
\usepackage{float}
\usepackage{multicol}
\usepackage{enumitem}

\usepackage{pgfplots}
\usepgfplotslibrary{colormaps}
\pgfplotsset{compat=1.17} 

\usepackage{amssymb}

\def\x{{\mathbf x}}
\def\L{{\cal L}}
\def\defeq{\vcentcolon=}

\DeclareMathOperator*{\argmin}{arg\,min}

\makeatletter
\def\blfootnote{\gdef\@thefnmark{}\@footnotetext}
\makeatother

\newcommand{\rebuttal}[1]{{#1}}

\title{Learning Continuous Representation of Audio \\ for Arbitrary Scale Super Resolution}

\name
 {Jaechang Kim$^1$$^*$, Yunjoo Lee$^1$$^*$, Seunghoon Hong$^2$, Jungseul Ok$^1$}
\address{$^1$Graduate School of AI, POSTECH \qquad  $^2$School of Computing, KAIST
    }

\begin{document}
%
\maketitle

\blfootnote{$^*$ equal contribution}
\blfootnote{This work was supported by NRF (No. 2021M3E5D2A01023887) and IITP ( No.2019-0-01906, Artificial Intelligence Graduate School Program(POSTECH)) grant funded by the Korea government (MSIT). 
Seunghoon Hong was supported by Samsung Electronics.}

\begin{abstract}
    Audio super resolution aims to predict the missing high resolution components of the low resolution audio signals. 
    While audio in nature is a continuous signal, current approaches treat it as discrete data (\emph{i.e.}, input is defined on discrete time domain), and consider the super resolution over a fixed scale factor (\emph{i.e.}, 
    it is required to train a new neural network to change output resolution).
    To obtain a continuous representation of audio
    and enable super resolution for arbitrary scale factor, 
    we propose 
    a method of implicit neural representation,
    coined
    Local Implicit representation for Super resolution of Arbitrary scale (LISA).
    Our method locally parameterizes a chunk of audio as a function of continuous time, and represents each chunk with the local latent codes of neighboring chunks so that the function can extrapolate the signal at any time coordinate, \emph{i.e.}, infinite resolution. 
    To learn a continuous representation for audio,
    we design a self-supervised learning strategy
    to practice super resolution tasks 
    up to the original resolution by stochastic selection.
    Our numerical evaluation shows 
    that LISA outperforms the previous fixed-scale methods 
    with a fraction of parameters, 
    but also is capable of 
    arbitrary scale super resolution
    even beyond the resolution of training data.

\end{abstract}
\section{Introduction}
\label{sec:intro}

    \rebuttal{
        Audio super resolution is a task to construct the missing high resolution components of given audio in low resolution.
        The problem is also known as bandwidth extension and sampling rate conversion (SRC), 
        and has been studied extensively in signal processing community.
        In early works~\cite{evangelista2003design, oppenheim99},
        super-resolution of factor $\frac{L}{M}$ is performed by a sequence of 
        $L$-fold upsampling interpolation, low-pass filtering, and then $M$-fold downsampling.
        %
        Note that the output quality of non-integer scale algorithm is mainly determined by upsampling method, e.g., spline~\cite{spline2012}.
        Classical approaches for 
        upsampling have been devised with
        a set of probabilistic models,
        e.g., 
        Gaussian mixture models~\cite{pulakka2011gmm}, hidden Markov models~\cite{jax2003markov} and linear predictive coding~\cite{bachhav2018predictivecoding}.
        %
        Recently,
        the advances of deep learning techniques
        have made
        remarkable improvement of upsampling,
        including but not limited to 
        U-Net architecture \cite{kuleshov2017superres},  
        Generative Adversarial Networks (GANs) \cite{eskimez2019, su2021bandwidth},
        and normalizing flows \cite{zhang2021wsrglow}.
}
    The existing deep learning methods however
    focus on the fixed scale super resolution,
    in which 
    neural network is used to have discrete representation 
    of audio signal only at a given set of coordinates corresponding to 
    the target resolution.
    However, audio is continuous in time, and this motivates us to 
    obtain a continuous representation of audio signal,
    which further enables the audio super resolution of arbitrary scale factor.
    Meanwhile, for the continuous representation of natural data such as 
    images and audio, implicit neural representation is an emerging area 
    where neural network is employed as a representation function 
    mapping from continuous coordinates to the corresponding signals (e.g., RGB value on a pixel in image, and amplitude on a sampling point in audio)
    ~\cite{sitzmann2019siren,part2019deepsdf,mildenhall2020nerf, chen2021liif}.
    %
    %
    
    We propose Local Implicit representation for Super resolution of Arbitrary scale (LISA) to obtain a continuous representation of audio.
    LISA consists of a pair of encoder and decoder.
    The encoder corresponds each chunk of audio
    to a latent code parameterizing a local input signal around the chunk.
    The decoder takes a continuous time coordinate
    and the neighboring set of latent codes around the coordinate, and predicts the value of signal at the coordinate.
    To train such a continuous representation for audio,
    we devise a self-supervised learning framework.
    Specifically, after downsampling the training data to the input resolution,
    we generate super resolution tasks
    of random scale factors up to the original resolution.
    As a training objective,
    we use a stochastic measure of audio discrepancy between
    the entire reconstructed and original signals in waveform and spectrogram,
    in which discrepancy closer to the target chunk receives
    higher random weight so that 
    the local latent code of a chunk
    captures a characteristic of the global audio signal, while
    focuses on the local signal around the chunk.

    Thanks to the continuous representation,
    LISA enables the arbitrary scale super resolution,
    by requesting the prediction of signal for the set of time coordinates
    corresponding to any arbitrary target resolution.
    In addition, the proposed method has several advantages in audio super resolution,
    compared to the existing works on implicit neural representation.
    In \cite{sitzmann2019siren}, 
    a demonstration of implicit neural representation for audio is provided. However, it requires a training procedure of neural network per signal,
    whereas our method does not. LISA with local implicit representation is particularly useful
    in audio super resolution since the encoding and decoding can be performed once
    a small chunk is obtained, i.e., low latency.
    The methods of \cite{mildenhall2020nerf,chen2021liif}
    are analog to ours in a sense of localizing implicit representations.
    However, they are dedicated to computer vision applications such as 3D vision \cite{mildenhall2020nerf} and 2D image \cite{chen2021liif}, while we focus on the local implicit representation of audio.
    Indeed, our approach includes some specialized features such as 
    the pair of neural encoder and decoder to process audio data and 
    the self-supervised learning with the audio discrepancy.
    In addition, we discover the advantages of having the local implicit representation
    in terms of the low latency and the arbitrary scale factor 
    as input audio is often streaming in audio super resolution,
    while such benefits are overshadowed in computer vision applications since 
    the vision data are given in a lump rather than a streaming fashion.
    We further describe differences in details in Section~\ref{sec:method}.

    \begin{figure}[!t]
        \includegraphics[width=\linewidth, trim={0cm, 0cm, 0cm, 0cm}]{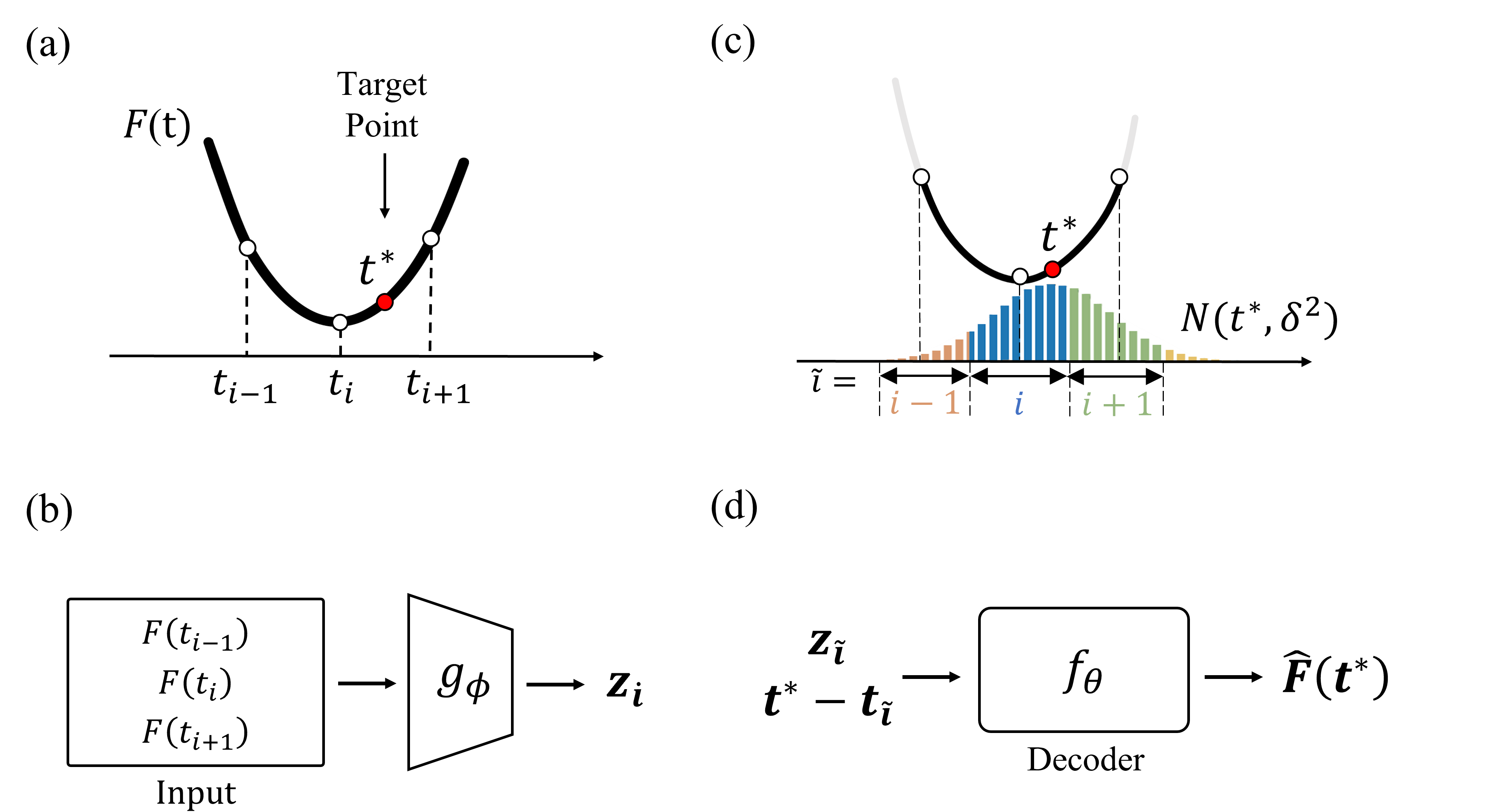}
        \caption{Model architecture.
        \rebuttal{
        {\bf (a)} The white circles denote sampled points of the input signal and the red circle denotes the point of interest.
        {\bf (b)} A convolutional encoder extracts local continuous representation.
        {\bf (c)} An index ${\tilde{i}}$ is selected stochastically from $\mathcal{N}(t^*, \delta^2)$.
        {\bf (d)} The amplitude of $t^*$ is predicted by the relative distance around $t_{\tilde{i}}$ and a latent code. 
        $z_{\tilde{i}}$ represents a continuous local signal.
        }
        }
        \label{fig:model}
        \vspace{-0.15in}
    \end{figure}

    \begin{figure}[t!]
        \includegraphics[width=.9\linewidth, trim={0cm, 1cm, 0cm, 0cm}]{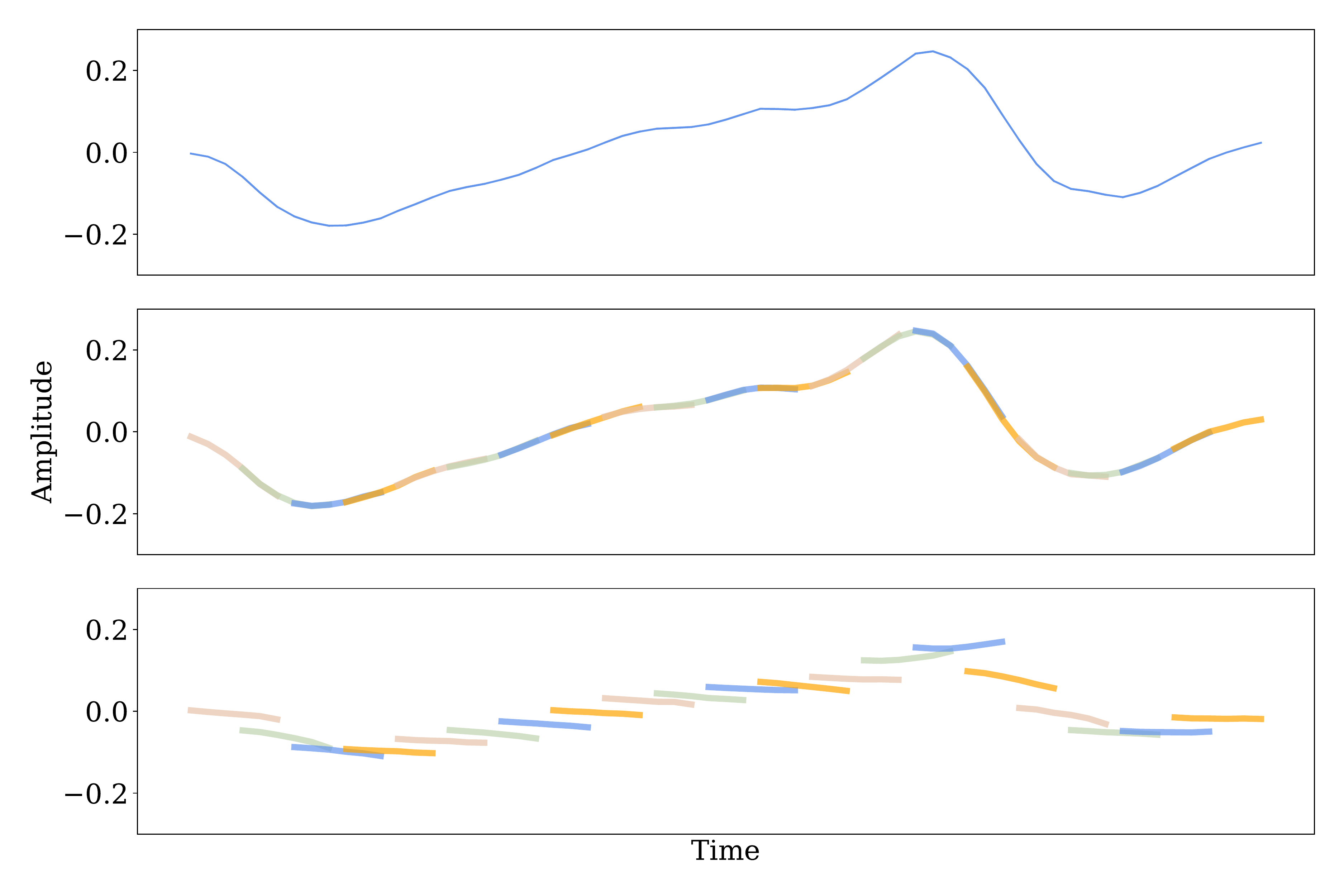}
        \caption{An example of local continuous representation. 
                {\bf(top)} target signal 
                {\bf(mid)} local signals trained with stochastic selection
                {\bf(bottom)} local signals trained with local ensemble.
                }
        \label{fig:local-signal}
        \vspace{-0.15in}
    \end{figure}

\section{Method}
\label{sec:method}


    Audio in nature is continuous, 
    denoted by $F(t)$ for continuous $t \in \mathbb{R}$, whereas
    we discretely observe $F(t)$ for $t = t_i$'s in every sampling period $\frac{1}{R_{\text{in}}}$,
    where $R_{\text{in}} > 0$ is the input resolution, and $t_i$ is the $i$-th temporal coordinate, i.e., $t_{i} - t_{i-1} = \frac{1}{R_{\text{in}}}$.
    Our method, LISA, aims to obtain  continuous representation $\hat F(t)$ for continuous $t$
    given a local part of discrete samples $F(t_i)$'s only around $t$ as input
    so that it provides the arbitrary scale super resolution with low latency.
    To do so,
    LISA employs encoder $g_\phi$ and decoder $f_\theta$
    with neural network parameters $\phi$ and $\theta$, respectively.
    We use encoder $g_\phi$ to extract
    local latent code $z_i$ from $(2k+1)$ samples around time~$t_i$,
    i.e.,
    \begin{align} \label{eq:latent_code}
    z_i := g_\phi(F(t_{i - k}), ..., F(t_{i + k})) \;.
    \end{align}
    Let $i(t)$ be the closest index to $t$, i.e., $i(t) :=  \argmin_i |t- t_i|$
    where tie breaks with preference to smaller $i$.
    We denote by $z(t) := (z_{i(t)-1}, z_{i(t)}, z_{i(t)+1})$ the set of local latent codes
    corresponding to $t$.
    Then, LISA represents the signal at any time $t \in \mathbb{R}$
    using decoder $f_\theta$ as follows:
    \begin{align} \label{eq:prediction}
    \hat{F}(t) := f_\theta (t - t_{i(t)}; z(t)) \approx {F}(t) \;.
    \end{align}
    When we want the super resolution from $R_{\text{in}}$ to $R_{\text{out}}$,
    the output signal is predicted by putting the sequence of time coordinates every $\frac{1}{R_{\text{out}}}$ to $\hat{F}(\cdot)$. 
    We note that ignoring computational cost for encoding and decoding,
    a theoretical upper bound of latency to predict signal in high resolution around $t$
    is only $\frac{k+1}{R_{\text{in}}}$.
    A graphical illustration in Fig~\ref{fig:model} summarizes LISA.

\subsection{Model Architecture}
\label{sec:architecture}

\noindent\textbf{Encoder. }
    To induce the temporal correlation in an audio signal, we use convolutional networks for encoder $g_\phi$,
    which produces a latent vector summarizing a few consecutive data points.
    Our fully convolutional encoder consists of 4 1D-convolution layers with the kernel size of \{7, 3, 3, 1\} and the channel size of \{16, 32, 64, 32\}.
    Note that the kernel size of convolutional network determines 
    the range of receptive field of encoder $g_\phi$ in \eqref{eq:latent_code}, 
    i.e., 
    in our choice, $k=5=3+1+1+0$ as $7=2\cdot 3 +1$, $3=2\cdot 1 +1$, and $1=2\cdot 0 +1$.
    Note that selecting larger $k$ may force to embed longer pattern inside signal,
    but it causes longer latency and requires a larger number of parameters.

\vspace{0.1in}
\noindent\textbf{Decoder. }    
    For decoder $f_\theta$, we use a 5-layer MLP with ReLU activation,
    of which input to predict signal at $t$
    is the concatenated vector of the relative coordinate $t - t_{i(t)}$
    and the set $z(t)$ of local latent codes around $t$.
    In \cite{sitzmann2019siren, mehta2021modulated},
    as a part of empowering representation of features in high frequency,
    periodic activations, such as sine function, have been proposed instead of ReLU.
    However, according to our experiment,
    such periodic activations make
    the training highly sensitive to initialization and hyperparameters.
    We hence choose to use ReLU for stable training,
    and small enough $k$ to express meticulous local representation. 
    The encoder generates the concatenated latent vector $z(t) = (z_{i(t)-1}, z_{i(t)}, z_{i(t)+1})$,
    which adds more information on the signal at a small computational cost.

\vspace{-0.11in}
\subsection{Training Strategy}
\label{sec:training}
To train encoder $g_\phi$ and decoder $f_\theta$, 
we generate a set of {\it self-supervised learning} tasks
from a training dataset of resolution $R_{\text{data}}$.
If we aim at super resolution from $R_{\text{in}}$, 
dataset resolution $R_{\text{data}}$ needs to be greater than $R_{\text{in}}$.
Each self-supervised learning task is
a super resolution task from $R_{\text{in}}$ to $R_{\text{out}}'$,
where $R_{\text{out}}'$ is drawn uniformly at random from the interval $[R_{\text{out}}^-, R_{\text{out}}^+]$,
and the self-supervision
can be obtained from downsampling the training dataset at resolution $R_{\text{in}}$
and $R_{\text{out}}'$
with sinc interpolation~\cite{shanon1949sinc}.
In Section~\ref{sec:exp}, we report experiment results from
several choices of $(R_{\text{in}}, R_{\text{out}}, R_{\text{out}}^-, R_{\text{out}}^+)$.
In what follows, 
we describe the remaining details about
training procedure with the generated task.

\vspace{-0.11in}
\subsubsection{Loss function}
\vspace{-0.05in}
    Given the generated super resolution task from $R_{\text{in}}$ to $R$,
    we let $T$ be the set of time coordinates corresponding to $R$,
    and $\x$ be the corresponding supervision, i.e., the sequence of $F(t)$
    for $t \in T$.
    Using encoder $g_\phi$ and decoder $f_\theta$,
    we generate the sequence of predictions for $t \in T$, denoted by $\hat{\x}$,
    with a random perturbation, which will be elaborated in Section~\ref{sec:subsampling}.
    We note that the perturbed prediction $\hat{\x}$ is used only for training
    and robustifies the prediction, 
    while for testing, the prediction is performed as described in \eqref{eq:prediction}.
    Using gradient-based optimizer with back-propagation, we train
    encoder $g_\phi$ and decoder $f_\theta$ to minimize
 a measure of discrepancy from the perturbed prediction $\hat{\x}$ to the supervision $\x$.
    To capture the discrepancy both in time and frequency domains,
    we measure it with L1 loss
    ${\cal L}_{\text{wave}}(\x, \hat \x) = \|\x - \hat \x \|_{1}$,
    and multi-scale spectrogram loss
    ${\cal L}_{\text{spec}}$ \cite{parallelwavegan}.
    Then, the training loss is given as:
        \begin{align} \label{eq:loss}
            {\L}( \x, \hat \x )
            = {\L}_{\text{wave}}( \x, \hat \x ) + \lambda {\L}_{\text{spec}}(\x, \hat \x ) \;, 
        \end{align}
        where
        with balancing hyperparameter $\lambda >0$.

\vspace{-0.05in}
\subsubsection{Stochastic selection of a local latent code}
\label{sec:subsampling}
\vspace{-0.05in}
We now describe how we generate the perturbed prediction $\hat\x$ in \eqref{eq:loss}. 
The perturbation is designed to reduce the disagreement
between two neighboring local predictions at the midway.
    In order to alleviate this issue, 
    \cite{chen2021liif} proposes an ensemble method by taking 
    the weighted sum of the predicted signals from different local predictions,
    where the weights are computed to be inversely proportional to geometric distance in coordinates.
    If we applied the ensemble method, then 
    the prediction at $t$ would be
    $
        {\frac{(t - t_{i})  f_\theta({z_{i+1}}, t_{i+1} - t) + (t_{i+1} - t)  f_\theta({z_{i}}, t - t_{i})}{t_{i+1} - t_i } }\;,
    $
    for $t \in [t_i, t_{i+1}]$. 
    However, we found that as shown in Fig~\ref{fig:local-signal},
    training with the ensemble make considerable errors in the local predictions,
    although the ensemble has small errors. 
    Hence, we rather choose an input latent vector randomly among a few closest latent vectors around the target coordinate.
    To do so, we perturb $i(t)$
    and use $ \label{eq:sto}
            \tilde{i}(t) \defeq i(t + \eta),$
    for training, 
        where $\eta$ is zero-mean Gaussian noise
        with variance $\delta^2 > 0$, i.e., $\eta \sim {\cal N}(0, \delta ^2)$.
        $\delta$ controls the desired covering range of a local signal.
        We use $\delta$ as the distance between $t_i$ and the border $\frac{t_i + t_{i+1} }{ 2}$.
    Through the stochastic selection of local latent codes, whichever one is chosen among the closest latent vectors, either one should be able to independently represent high resolution at the target coordinate. Therefore, each latent vector is guided to depict the details of the local signal to predict the high resolution component without referring to predictions from the other closest latent vectors.
    By sampling the latent vectors, 
    a latent vector $z_i$ also learns to represent data out of the interval  
    $ [ \frac{t_{i-1} + t_i}{2} , \frac{t_{i} + t_{i+1}}{2} ] $.
    Fig~\ref{fig:local-signal} shows the local signals are connected to adjacent local signals.

\vspace{-0.1in}
\section{Experiments}
\label{sec:exp}
\vspace{-0.1in}
    \begin{figure*}[t]  
        \captionsetup[subfigure]{margin={0.9cm,0cm}}
        \subfloat[Ground Truth]{
        	\begin{minipage}[c][0.6\width]{
        	   0.32\textwidth}
        	   \centering
        	   
        	   \includegraphics[width=1.1\linewidth, trim={0cm, 0cm, 0cm, 1cm},
        	   ]{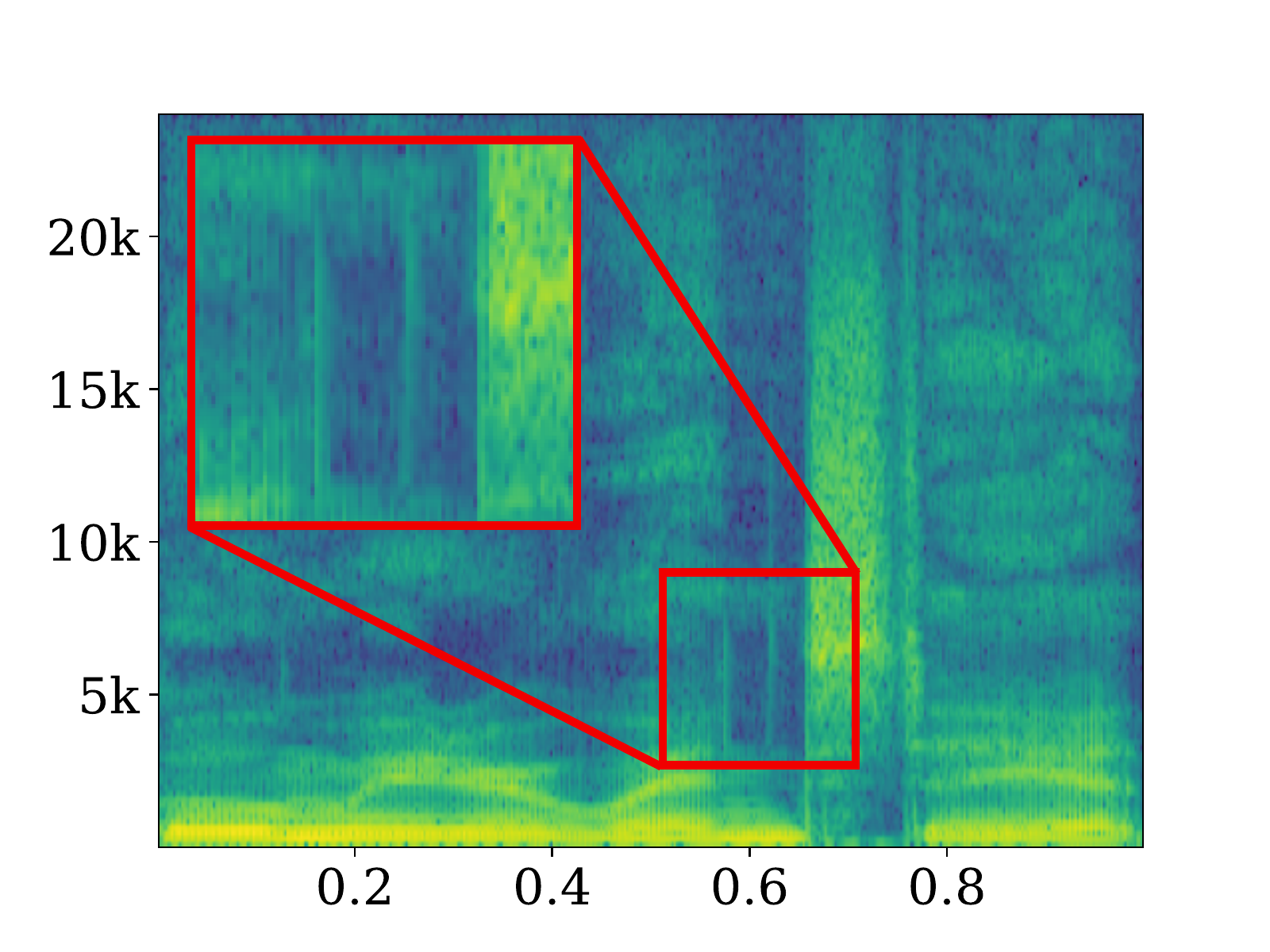} 
        	\end{minipage} }
        \subfloat[LISA (ours)]{
        	\begin{minipage}[c][0.6\width]{
        	   0.32\textwidth}
        	   \centering
        	   \includegraphics[width=1.1\linewidth, trim={0cm, 0cm, 0cm, 1cm}, ]{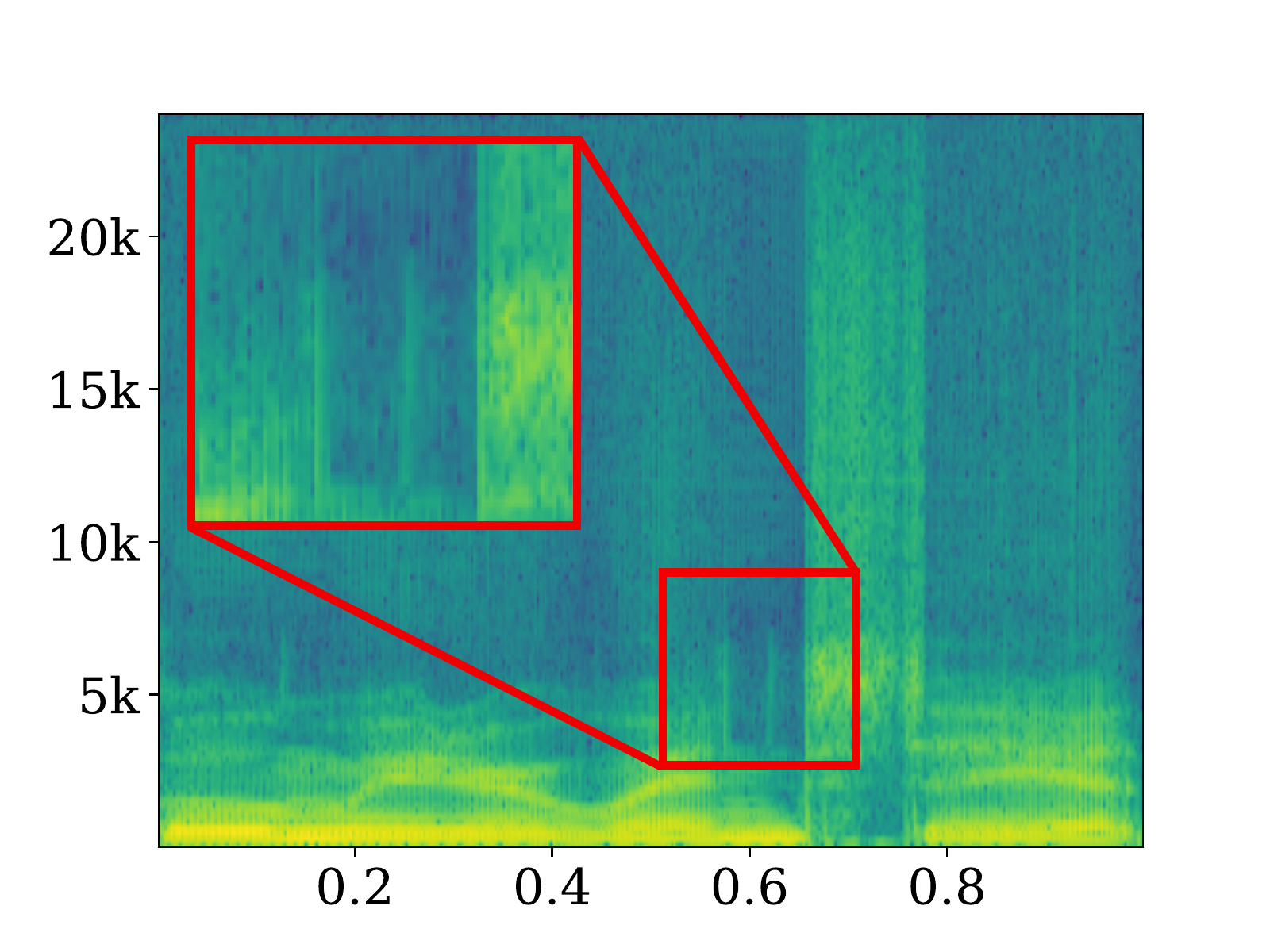}
        	\end{minipage} }
        \subfloat[WSRGlow \cite{zhang2021wsrglow}]{
        	\begin{minipage}[c][0.6\width]{
        	   0.32\textwidth}
        	   \centering
        	   \includegraphics[width=1.1\linewidth, trim={0cm, 0cm, 0cm, 1cm},  ]{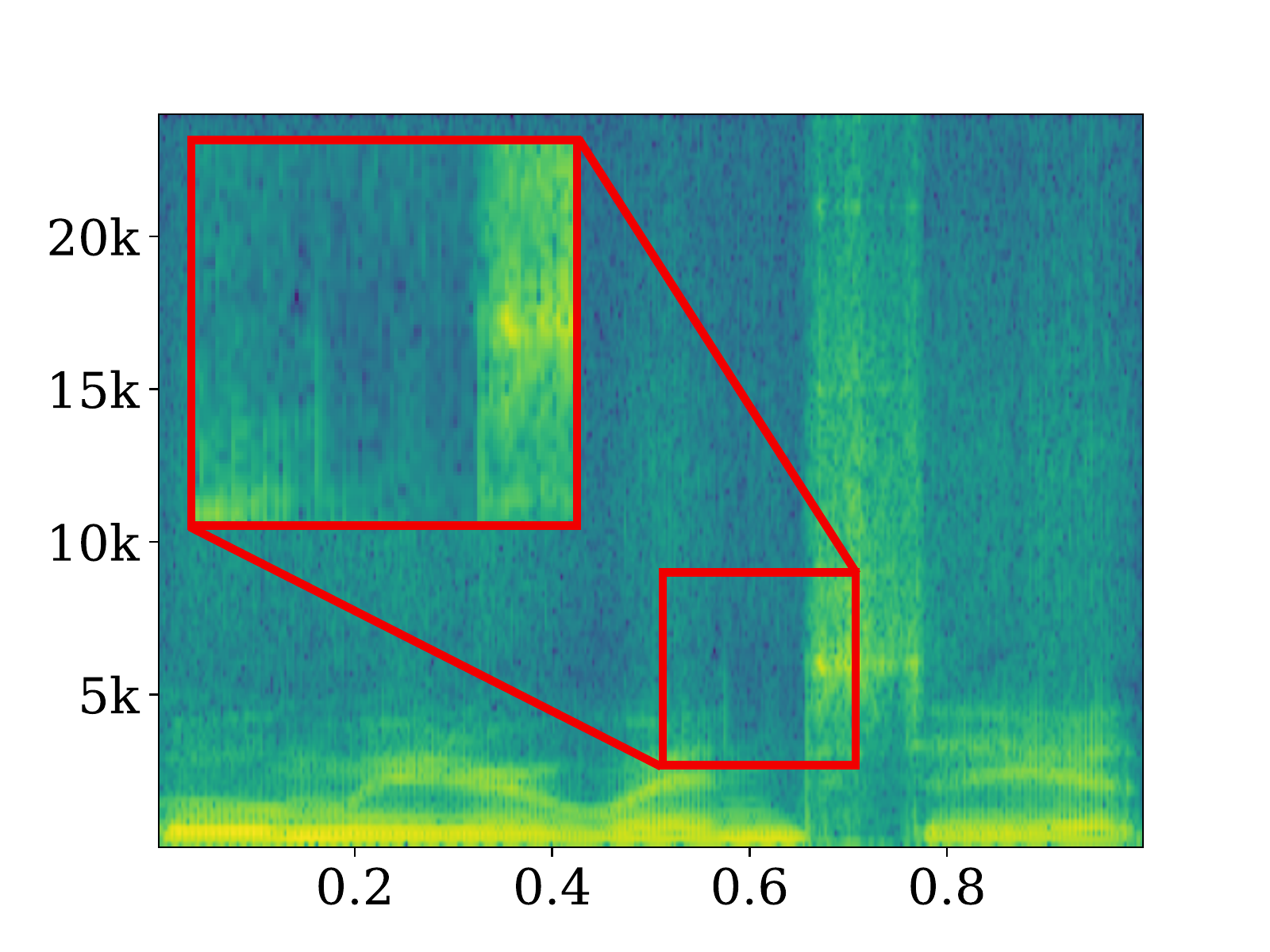}
        	\end{minipage} }
        \vspace{-0.1in}
        \caption{Super resolution results in spectrogram. 
        (a) Ground Truth is 48kHz signal,
        (b) and (c) are super resolution results of the $\times 4$ (12kHz $\to$ 48kHz) setting in Table~\ref{tab:comparison}.
                }
        \label{fig:spec}
        \vspace{-0.2in}
    \end{figure*}
    


    Throughout the experiments,
    we use CSTR VCTK corpus~\cite{vctk}, 
    a high-quality speech dataset in 48kHz 
    uttered by 109 English speakers.
    We inherit the same multi-speaker setting used in \cite{kuleshov2017superres},
    where the first 99 speakers are used for training and the rest are served as test set.
    We compare our method with the existing audio super resolution methods including AudioUNet~\cite{kuleshov2017superres}, AudioTFilm~\cite{birnbaum2019tfilm}, and WSRGlow~\cite{zhang2021wsrglow}. 
    To be as fair as possible, we use the pre-trained models of baselines if available, and 
    we reproduce using the original authors' official source codes otherwise.
    We evaluate the performance of super resolution in terms of Signal-to-Noise-Ratio (SNR) and Log-Spectral-Distance (LSD).
    SNR is defined as
    \begin{equation}
        \text{SNR}(\x,\hat \x)=10\log \frac{\|\x\|_2^2}{\|\x - \hat \x\|_2^2} \;,
    \end{equation}
    where $\hat x$ is the predicted signal and $x$ is the ground truth signal.
    LSD is defined as 
    \begin{equation}
        \text{LSD}(\x, \hat \x)=\frac{1}{L}\sum_{l=1}^L\sqrt{\frac{1}{K}\sum_{k=1}^K \left(X(l,k)-\hat X(l,k)\right)^2} \;, \label{eq:LSD}
    \end{equation}
    where, using
    Short-Time Fourier Transform
    ($\text{STFT}$),
    $X (l,k) = \log | \text{STFT}^{(l,k)}(\x) | ^ 2 $ is log-spectral power magnitudes of $\x$ with index frame $l$ and frequency $k$, and $\hat X$ is defined similarly but for $\hat \x$.
    We use Adam optimizer \cite{adam} with learning rate of 0.001.
    We train the model for 50 epochs, and decay the learning rate by half at every 5 epochs. 
    We use gradient clipping with $0.001$ max norm for stable training.
    
\subsection{Fixed scale super resolution}
\label{sec:fixed}
    \begin{table}[!t]
    \vspace{0.1in}
        \footnotesize
        \centering
        \caption{Objective comparison with other methods.}
        \vspace{-0.1in}
        \begin{tabular}{c|rr|rr|r}
        \toprule
        \multicolumn{1}{l}{} & \multicolumn{2}{|c|}{$\times2$ (24k$\to$48k)}  & \multicolumn{2}{c}{$\times$4 (12k$\to$48k)} & \multicolumn{1}{|c}{\# params} \\
        \multicolumn{1}{r|}{} & \multicolumn{1}{l}{SNR$\uparrow$} & \multicolumn{1}{l|}{LSD$\downarrow$}  & \multicolumn{1}{l}{SNR$\uparrow$} & \multicolumn{1}{l|}{LSD$\downarrow$} \\ \midrule
        AudioUNet \cite{kuleshov2017superres}            & 21.68                   & 1.31           & 18.55                   & 2.11                & 71M    \\
        AudioTFilm \cite{birnbaum2019tfilm}            & 22.23                   & 1.05           & 19.51                   & 2.02               & 68M     \\
        WSRGlow \cite{zhang2021wsrglow}        & 25.29          & 0.61        & 19.41          & 1.01     & 229M      \\ 
        LISA (Ours)         & \bf{30.7}        & \bf{0.58}         & \bf{24.16}          & \bf{0.81}     & 89k     \\ \bottomrule
        \end{tabular}
        \label{tab:comparison}
    \end{table}

    We train our model, LISA, on two tasks of fixed scale resolution: i) from 12kHz to 48kHz; and ii) from 24kHz to 48kHz.
    Table~\ref{tab:comparison} compares LISA to the baselines,
    in which LISA shows the best performance in terms of both SNR and LSD.
    It is remarkable that in terms of the number of model parameters,
    our model is at least 764 times lighter than the others.
        In Fig~\ref{fig:spec}, 
        LISA captures detailed features in the spectrogram, which are highlighted with red boxes.
        However, WSRGlow misses the details despite it uses 2,573 times more parameters. 
        The difference is noticeable in listening.

\subsection{Arbitrary scale super resolution}
\label{sec:exp-arbitrary}

    \begin{table}[t!]
        \footnotesize
        \centering
        \caption{Evaluation of arbitrary scale super resolution and ablation study.
                The same model is used for 2$\times$, 3$\times$, and 6$\times$ super resolution.
                -sto, -spec denotes models trained without stochastic selection and without multi-scale spectrogram loss.
        }
        \label{tab:arbitrary}
        \vspace{-0.1in}
        \begin{adjustbox}{width=\linewidth}
        \begin{tabular}{c|rr|rr|rr}
        \toprule
        \multicolumn{1}{l}{} & \multicolumn{4}{|c|}{in-distribution}         & \multicolumn{2}{c}{out-of-distribution}                        \\
        \multicolumn{1}{l}{} & \multicolumn{2}{|c|}{$\times$2 (8k$\to$16k)} & \multicolumn{2}{c|}{$\times$3 (8k$\to$24k)}      & \multicolumn{2}{c}{$\times$6 (8k$\to$48k)}                        \\
        \multicolumn{1}{l}{} & \multicolumn{1}{|l}{SNR$\uparrow$} & \multicolumn{1}{l|}{LSD$\downarrow$} & \multicolumn{1}{l}{SNR$\uparrow$} & \multicolumn{1}{l|}{LSD$\downarrow$} & \multicolumn{1}{l}{SNR$\uparrow$} & \multicolumn{1}{l}{LSD$\downarrow$}                      \\ \midrule
        
        LISA         & 25.66 & 0.94  & 24.15 & 1.17  & 22.17 & 1.23           \\  \midrule
        LISA (-sto)         & -0.18 & +0.01  & -0.05 & +0.00  & -0.04 & +0.01           \\  
        LISA (-spec)         & -0.08 & +0.00  & -0.07 & +0.00  & -0.04 & +0.01           \\  \bottomrule
        \end{tabular}
        \end{adjustbox}
    \end{table}

    To demonstrate the arbitrary scale super resolution using LISA,
    we obtain a single model with the self-supervised learning tasks 
    from $R_{\text{in}} = 8\text{kHz}$ to the random target resolution drawn from 
        $[R^-_{\text{out}}, R^+_{\text{out}}]=[8\text{kHz}, 24\text{kHz}]$,
        i.e., the training never uses
        any sound of 48 kHz resolution.
    
    \vspace{0.1in}
    \noindent\textbf{Out-of-distribution.} 
    In Table~\ref{tab:arbitrary}, the model is evaluated with 2$\times$, 3$\times$, and 6$\times$ super resolution tasks.
    In the out-of-distribution task, which our model upsamples with a scale of 6, our model even outperforms the results of 4x super resolution task of baselines. 
    Note that the 48kHz data, which is the output resolution of 6$\times$ setting, is never seen to the model while training.

    \begin{figure}[t!]
    \begin{tikzpicture}
    \begin{axis}[
        xmin=7000, xmax=41000,
        xtick={8000, 16000, 24000, 32000, 40000},
        xticklabels={
            $\times$1, $\times$2, $\times$3, $\times$4, $\times$5
            },
        scaled x ticks=false,
        ylabel=SNR (dB),
        width=\linewidth,
        height=3.9cm,
    ]
    \definecolor{color1}{HTML}{8080E0}
    \addplot[color=color1, mark=*]
        coordinates {
            (8000	,36.2061 )
            (10000,	30.6415 )
            (12000,	27.8758 )
            (14000,	26.3908 )
            (16000,	25.5308 )
            (18000,	25.0101 )
            (20000,	24.5984 )
            (22000,	24.3032 )
            (24000,	24.0519 )
            (26000,	23.82 )
            (28000,	23.6133 )
            (30000,	23.4275 )
            (32000,	23.2539 )
            (34000,	23.0984 )
            (36000,	22.9579 )
            (38000,	22.8201 )
            (40000,	22.6908 )
        };
    \end{axis}
    \end{tikzpicture}
    \vspace{-0.1in} 
    \caption{Super resolution performances at different scales. 
    }
    \label{fig:non-integer}
    \vspace{-0.2in} 
    \end{figure}
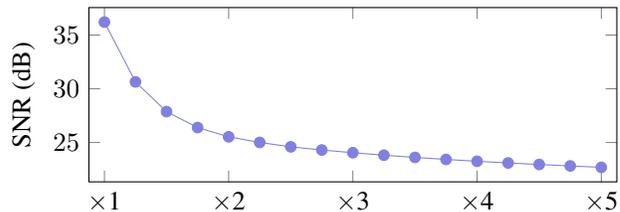
    
    \vspace{0.1in}
    \noindent\textbf{Non-integer scale SR. }    
    Also, we evaluated our results in non-integer scale.
    In Fig~\ref{fig:non-integer}, non-integer scale super resolution performance is no worse than  integer scale super resolution, does not show any fluctuation.
    The test data is generated with sinc interpolation from 48k data.


\smallskip
\vspace{0.1in}
\noindent\textbf{Ablation Study.}
In Table~\ref{tab:arbitrary}, we also assess the effect of each component through an ablation study.
    The perturbed prediction \eqref{eq:sto} in training is twice beneficial in terms of SNR gain
    compared to the spectrogram loss function in \eqref{eq:loss} in the 2$\times$ super resolution.
    In addition, 
    Fig~\ref{fig:local-signal} shows that 
    the discontinuity between the local signals is indeed regulated by
    the perturbed prediction \eqref{eq:sto}.

\section{Conclusion}
We propose LISA to obtain a continuous representation of an audio
as a neural function, which enables
us to perform the arbitrary scale audio super resolution.
To train LISA, we devise a sophisticated self-supervised learning strategy,
equipped with the perturbed prediction in training
and the loss function for audio signal.
Our experiment shows that LISA outperforms the other baselines 
in terms of super resolution performance with a remarkably small size.
We expect that our model
has a great potential for online audio super resolution applications
thanks to the light size compared to other baseline models
but also the low latency from the local implicit representation.

\vfill\pagebreak

\bibliographystyle{IEEEbib}
\bibliography{refs}

\end{document}